\documentclass[12pt]{article}
\usepackage{graphicx}
\newcommand{\be}{\begin{equation}}
\newcommand{\bea}{\begin{eqnarray} \nonumber}
\newcommand{\ee}{\end{equation}}
\newcommand{\eea}{\end{eqnarray}}

 \def\(({\left(}
 \def\)){\right)}
\def\[[{\left[}
\def\]]{\right]}

\def \form#1 {eq. (\ref{#1}) }
\def \parziale#1#2  {{\partial {#1} \over \partial {#2}}}

\def \ba#1 {\overline{#1}}

\def  \eps{\epsilon}
\def  \s {\vec{s}}
\begin{document}

\title{Some remarks on  the survey decimation algorithm for K-satisfiability}
\author{ Giorgio Parisi\\
Dipartimento di Fisica, Sezione INFN, SMC and
 UdRm1 of INFM,\\
Universit\`a di Roma ``La Sapienza'', \\
Piazzale Aldo Moro 2,
I-00185 Rome (Italy)}

\maketitle

\begin{abstract}

\noindent In this note we study the convergence of the survey decimation algorithm. An analytic formula for the 
reduction of the complexity during the decimation is derived. The limit of the converge of the algorithm are estimated 
in the random case: interesting phenomena appear near the boundary of convergence.
\end{abstract}

\section{Introduction}

Recently a very powerful algorithm has been proposed \cite{MPZ,MZ} for finding the solution of the random 
K-satisfiability problem  \cite{KS,sat,sat0}.  This new algorithm (see also \cite{BMWZ,BMZ,MPWZ,P1,P2}) is based 
on the survey-propagation equations that generalize the older approach based on the ``Min-Sum''\footnote{The ``Min-Sum'' 
is the the zero temperature limit of the ``Sum-Product'' algorithm and sometimes is also called belief propagation.  In 
the statistical mechanics language \cite{MPV} the belief propagation equations are the extension of the TAP equations 
for spin glasses \cite{TAP} and the survey-propagation equations are the TAP equations generalized to the broken replica 
case.} algorithm \cite{BP,MPV,factor,MoZ} .

The aim of this note is to progress in the understanding of the deep reasons of the very good performance of this survey 
decimation algorithm.  In the second section of this note we present a fast heuristic derivation of the survey equations.  In the 
third section we analyze the decimation algorithm and we give an analytic formula for the decrease in the complexity during 
the decimation.  Finally, in the fourth section we present some numerical studies of the decimation algorithm in 
the random case: they suggest  an upper bound on the region where the algorithm may converge; we 
notice the appearance of new phenomena near the boundary.

\section{A fast heuristic derivation of the survey equations}
\subsection{The random K-sat problem }
In the random K-sat problem there are $N$ variable $\sigma(i)$ that may be true of false (the index $i$ will sometime 
called a node).  An instance of the problem is given by a set of $M\equiv \alpha N$ s.  For $K=3$ each clause is 
characterized by a set of three nodes ($i_{1}$,$i_{2}$, $i_{3}$), that belong to the interval $1-N$ and by three Boolean 
variables ($\beta_{1}$,$\beta_{2}$, $\beta_{3}$).  In the random case the $i$ and $b$ variables are random with flat 
probability distribution.  Each clause $c$ is true if the expression
\be
E_{c}\equiv (\sigma(i^{c}_{1})\ XOR\ \beta^{c}_{1}) \ OR \ (\sigma(i^{c}_{2})\ XOR\ \beta^{c}_{2}) \ OR \ 
(\sigma(i^{c}_{3})\ XOR\ \beta^{c}_{3})
\ee
is true.
The problem is satisfiable iff we can find a set  of the variables $\sigma$ such that all the clauses are true. The 
entropy \cite{MoZ} of a 
satisfiable problem is the logarithm of the number of the different sets of  the $\sigma$  variables that make all the clauses  true.

To a given problem we can associate a graph (the factor graph \cite{factor}) where the nodes are connected to the 
clauses ($3\alpha$ in average) and each clause is connected to three nodes.  The properties of this graph play a 
very important role. Some of the considerations we are going to use in the following we be valid in the random case, 
where when $N\to \infty$ at fixed $\alpha$ the factor graph is locally a tree.

\subsection{Beliefs, warning and surveys}

Generally speaking if the problem is satisfiable, very often there are many configurations of the boolean variables that 
satisfy it. One would like to have some description of the set of configurations that satisfy the all the clauses (in the 
rest of this paper 
we will call the configurations that satisfies all the clauses  \emph{legal configurations}).

In the simplest approach one introduces the strong belief or warning variable $b(i)$. They may takes tree values: true, false or unknown 
(in the context of colorability \cite{MPWZ} we can introduce a new color: white \cite{P2,BMWZ}).

In an heuristic approach one assumes that for certain values of the parameters the set of legal configurations may be 
decomposed into sets $C_{\gamma}$ such that each element of the set is \emph{near} to the other elements of the set and is 
far the elements of the \emph{other} sets.

For each set we define the warning \footnote{The usual beliefs at the node $i$ is a variable $p_{\gamma}(i)$ that 
represent the probability that the variable $\sigma(i)$ is true in a randomly chosen legal configurations of the set 
$C_{\gamma}$.  Obviously the warning $b_{\gamma}(i)$ is true if $p_{\gamma}(i)=1$, is false if $p_{\gamma}(i)=0$ and is unknown if 
$0<p_{\gamma}(i)<1$.} corresponding to a given set according to the following rule:
\begin{itemize}
    \item If $\sigma(i)$ is true in all the legal configurations of the set $C_{\gamma}$, $b_{\gamma}(i)$ is true.
    \item If $\sigma(i)$ is false in all the legal configurations of the set $C_{\gamma}$, $b_{\gamma}(i)$ is false.
    \item If $\sigma(i)$ is true in some legal configurations  of the set $C_{\gamma}$ and it is false in some legal configurations of 
    the same set, $b_{\gamma}(i)$ is unknown (or indifferent).
\end{itemize}
One can also introduce directional warning ($b_{\gamma}(i,c)$): they are defined to be the strong beliefs at $i$ 
in absence of the clause $c$ (we consider only the case where $i \in c$).

Using this definition of warning it can be argued that in the limit $N \to \infty$ for a random problem the directional 
warnings satisfy (or quasisatisify) the warning propagations equations (that we will write later).  It is possible to argue that we 
can associate to any legal configuration a solution (or a quasi-solution) of the warning propagations equations, so that 
the legal configurations can be divided into clusters according to the solution of the warning propagation 
equations they correspond to \cite{P2}.

A very important quantity is the number of  solutions of the warning equations that correspond to some legal 
configuration (such a belief will be called a legal warning).  This number is given by $\exp(\Sigma)$, where 
$\Sigma$ is called the complexity of the problem.

We would like to compute the complexity $\Sigma$ and get some information on the structure of the warnings.  It 
argued that this can be done in the following way \cite{MPZ,MZ}.  One introduces the survey $\s(i)\equiv 
(s_{T}(i),s_{I}(i),s_{F}(i))$ that is a a three component vector: the probability that in the set of legal strong 
beliefs $b_{\gamma}(i)$ is true, indifferent and false is given by $s_{T}(i)$, $s_{I}(i)$ and $s_{F}(i)$ respectively.  
In a similar way we introduce the directional survey $\s(i,c)$ that is the survey at $I$ with the clause $c$ removed.  
Obviously a survey satisfies a normalization condition:
\be
s_{T}+ s_{I}+ s_{F}=1 \label{NORM} \ .
\ee

It can be argued that for large $N$ the complexity can be approximatively written as \cite{MPZ,MZ,MP1,MP2}.
\be
\Sigma= \sum_{i=1,N} \Sigma_{N}(i) -\sum_{c=1,M}(K(c)-1) \Sigma_{C}(c)
\ee
where $K(c)$ is the number of boolean variables that enters in the clause $c$. 
We have also defined 
\be
\Sigma_{C}(c)=\ln\((  (1- \prod_{a=1,K(c)} (\beta_{a}^{c}¥ \cdot \s(i_{a},c) )_{F} \)) \ ,
\ee
where the product of a boolean variable $\beta$ with a survey $\s$ is $\s$ itself if $\beta$ is true and it is given by the 
 vector  $s_{F},s_{I},s_{T}$ if the variable $\beta$ is false ($\s_{F}$ denotes the third components of the vector $\s$).

The definition of $\Sigma_{C}(i)$ is slightly more involved. It is given by
\be
\Sigma_{N}(i)=\ln \(( \left| \prod_{c\in i} \vec{u}(i,c) \right| \)) \ ,
\ee
where $\vec{u}(i,c)$ is a message from a clause to a node and we have defined the product of two vectors in the 
following way
\be
\vec{v}\vec{w}=\{v_T \ w_T+v_{I}\  w_T  +v_T\ w_{I}\ , \ v_{I} \ w_{I}\ ,  \ v_F\ w_F+v_{I}\  w_F  +v_F\  w_{I} \}
		 . \label{B}
\ee
The vector $(0,1,0)$ is the identity. The norm $|\vec{v}|$ of the vector $\vec{a}$ is  defined by
\be
|\vec{v}|=v_{T}+v_{I}+v_{F}\ .
\ee
Surveys have norm 1.

We still have to define the message  from a clause to node ($\vec{u}(i,c)$). It is a normalized vector $|\vec{u}(i,c)|=1$
fixed  by the following condition :
\be
\Sigma_{C}(c)  =\ln |\vec{u}(i,c) \vec{s}(i,c)| \ ,
\ee 
where $|\vec{u}(i,c)|$ does not depend on $\vec{s}(i,c)$.  In the case where all the $b$ variables are true and  an explicit 
computation gives 
\be
\vec{u}(i,c)=(f,1-f,0)\ , \ \  \ \ \  f=\prod_{a=1,K(c),i_{a}\ne i} \s(i_{a},c) )_{F} \ .
\ee
The quantity $\Sigma_{N}(i)$ and $\Sigma_{C}(c)$ have the meaning of the variation of  the complexity when we add  the 
node $i$ and the clause $c$ respectively.

Heuristically one suppose that the surveys satisfy the  survey propagation equations that are defined to be the 
stationary equation of the complexity.
\be
{\partial \Sigma \over \partial \s(i_{a},c)} =0:
\ee
They can be written in an more explicit form as
\be
 \s(i_{a},c) ={\prod_{d\in i, d\ne c} \vec{u}(i,d) \over |\prod_{d\in i, d\ne c}\vec{u}(i,d) |} \label{EQ} \ .
 \ee
 The survey $\s(i)$ is given by the relation
 \be
 \s(i)\propto \prod_{d\in i,} \vec{u}(i,d) \propto \s(i,c)\vec{u}(i,c)  \ \ \    \forall c \in i. \ .
 \ee
 
 It is interesting that the warning equations have exactly the same form of the survey equations if we assume that 
 the surveys may be only one of the following three forms: $(1,0,0)$, $(0,1,0)$ and $(0,0,1)$.  \ Generally speaking we will only 
 consider in the following solutions of the surveys equations that are not of the previous form.
 
All this is heuristical. Independently of the derivation of the survey equations its interesting to study their 
properties on a random lattice. Numerical experiments and analytic computations \cite{MPZ, P1} suggest that in the limit $N \to 
\infty$ the survey equations have an unique non-trivial solutions (i.e. different from the trivial solution $s(i_{a},c)=\vec{I}$)
in the interval $\alpha_{L}< \alpha< \alpha_{U}$ 
($\alpha_{L}$ and $\alpha_{U}$ are are near to 3.91 and 4.36) and this unique solution may be obtained by iterations. 
In this interval the complexity is a decreasing function of $\alpha$ that changes sign at $\alpha^{*}\approx 4.267$.

The interval $\alpha_{L}< \alpha^{*}$ is interesting because here simple  methods have difficulties in find a legal 
configuration.
For a given problem in the interesting  region we can find solutions to the survey equations. This solution carries 
information on the on the legal configurations so that it is natural to try to use them in an algorithm to find  legal configurations.

\section{Survey inspired algorithm}
The basic hypothesis beyond the survey decimation algorithm is that the solution of the survey equations give reliable 
information on the problem. In particular we assume that:
\begin{itemize}
    \item If the complexity is positive, there exist legal configurations and the problem is satisfiable; if the complexity 
    is negative, there are no legal configurations and the problem is not satisfiable.
    \item If the survey equations have only the trivial solution ($s(i_{a},c)=\vec{I}$), the problem is easy and it can be 
    easily solved.  On the other hand if the survey equations have a non trivial solution with positive complexity, the 
    problem  has solutions but they may be difficult to be found.
 \end{itemize}
 
 Here we are taking for granted these two hypothesis. Our aim is to simplify the problem in such a way that it becomes an 
 easy problem. This will be done using the decimation algorithm introduced in \cite{MPZ,MZ} and described below.
 \begin{figure} \begin{center}    
       \includegraphics[width=0.70\textwidth]{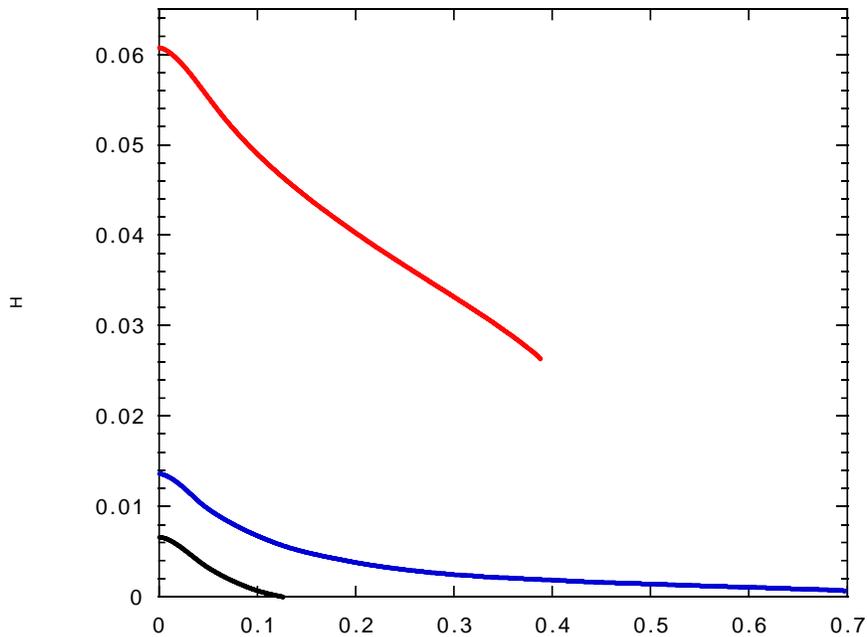}
      \end{center} \caption{ \label{EVOLVI}
 The complexity density $\Sigma$ as function of the fraction of decimated nodes in the region of positive complexity 
 for a problem with $N=3\ 10^{5}$
 for three values of $\alpha$ (i.e. 4.2. 4.25, 4.26 from above to below). 
}\end{figure}

\subsection{Decimation}
If a survey ($\s_{i}$) is very near to $(1,0,0)$ (or to $(0,0,1)$) in most of the legal solutions of the warning equations 
(and consequently in the legal configurations) the corresponding local variables will be true (or false).

The main step in the decimation procedure consists is starting form a problem with $N$ variables and to consider a 
problem with $N-1$ variables where $\s(i)$ is fixed to be true (or false).  We denote
\be
\Delta(i)=\Sigma^{N}-\Sigma^{N-1} \ .
\ee
If $\Delta(i)$ is small, the second problem it is easier to solve: it has nearly the same number of solutions of the warning equations 
an one variable less. (We assume that the complexity can be computed by solving the survey 
equations.)

The decimation algorithm proceeds as follows.  We reduces by one the number of variables choosing the node $i$ in the 
appropriate way, e.g. by choosing the variable with minimal $\Delta(i)$.  We recompute the solutions of the survey 
equations and we reduce again the number of variables.  At the end of the day two things may happen: 
\begin{itemize}
\item We arrive to a 
negative complexity (in this case we are lost),
\item The non trivial solution of the survey equation disappears.  If this 
happens the reduced problem is now easy to be solved  \footnote{It is also possible that the surveys propagation 
equations do not have anymore a solution that can be reached by iterations (e.g. if the denominator in eq.  \ref{EQ} is 
zero: apparently this happens only in the negative complexity region.}. 
\end{itemize}

This program may be successfully if we have a good criterion for choosing the point $i$ and estimating $\Delta(i)$.
Intuitively one would expect that if $\s(i)=(s_T,s_{I},s_{F})$ and $s_{T}$ is large we have that
\be
\Delta(i)=-\ln(1-s_{F})\ .
\ee
Indeed if we fix the variables $\sigma(i)$ to be true, we loose all the solutions of the warning equations such that 
$b(i)$ is false and therefore the total  number of solutions of the warning equations should decrease by a factor $(1-s_{F})$.
\begin{figure} \begin{center}    
      \includegraphics[width=0.70\textwidth]{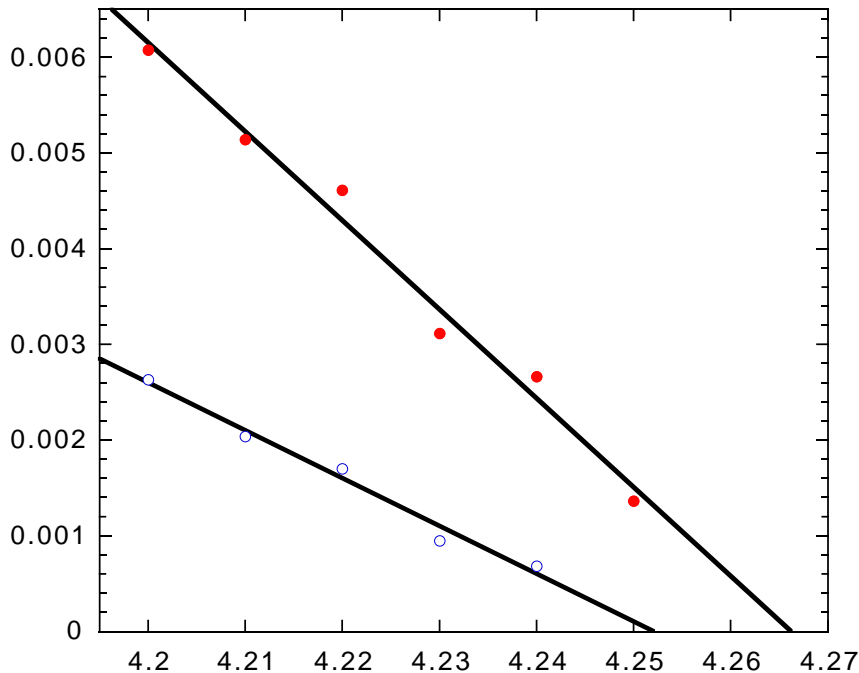}
      \end{center} \caption{ \label{FINAL}
 The initial (filled circles) and the final (empty circles) complexity density for a problem with $N=3\ 10^{5}$ as 
 function of $\alpha$.  The lines are a linear fit.}
\end{figure}

We now want to prove that this intuitive argument is correct. More precisely let us define the \textit{certitude} of  a 
survey as
\be
 s_{C} = \max(s_{F}+s_{I},s_{T}+s_{I})=1 - \min(s_{T},s_{F})\  .
\ee
If $s_C(i,c)=1-\epsilon$, we have that
\be
\Delta(i)=\ln(1-s_{C}(i))+O(\epsilon^{2})\ .
\ee
The crucial step consists in observing that the survey equations of the system with $N-1$ variables are the same of 
those of $N$ variables if we do not use the equations for $\s(i,c)$ and we set $\s(i,c)=(1,0,0)\equiv \vec{T}$.  Let us 
call $\s^{*}$ the solutions to these new survey equations.  In general $\s^{*}-\s=O(\epsilon)$.  The stationary 
equations imply that $\Sigma^{N}(\s)-\Sigma^{N}(\s^{*})=O(\epsilon^{2})$,.  At the end of the computation,we find that 
neglecting terms of order $\epsilon^{2}$ we have:
\be
\Delta(i)=\Sigma_{N}(i) -\sum_{c\in i}\ln(\vec{T} \vec{u}(i,c))  \ .
\ee
A detailed computations shows that the r.h.s. of the previous equation is just given by $\ln(1-s_{C})$.

If we neglect the terms of order $\eps^{2}$, the previous argument suggests the that the best variable to be eliminated 
are those that have the higher certitude, or the smallest value $\min(s_{T},s_{F})$.  This last criterion is similar, 
although different to the one used in \cite{MZ}, where the sites with maximum polarization, i.e. $|s_{T}-s_{F}|$.  The 
two quantities are obviously correlated: if the polarization is near to one also the certitude is near to one.

\section{The limits of the algorithm.}

Here in order to illustrate how the algorithm works we report for completeness the results of a few numerical 
experiments we have done on large samples (from $N=10^{4}$ to $N=3 \ 10^{5}$ near $\alpha^{*}$ where (for fastening the 
algorithm) a fraction $f=10^{-4}$ of the total variables has been decimated simultaneously.  In fig.  (\ref{EVOLVI}) for 
one sample with $N=3 \ 10^{5}$ we plot the complexity as function of the number of iterations for three different values 
of $\alpha$ where we have blocked the surveys with maximal polarization. We see that for the low value of $\alpha$ the 
method does work, the complexity jumps to zero coming from a positive value, while for the high value of $\alpha$ the 
complexity becomes negative.  A very similar is obtained is done in the case where we select the surveys using the 
maximum value of the \emph{certitude}.

\begin{figure} \begin{center}    
      \includegraphics[width=0.7\textwidth]{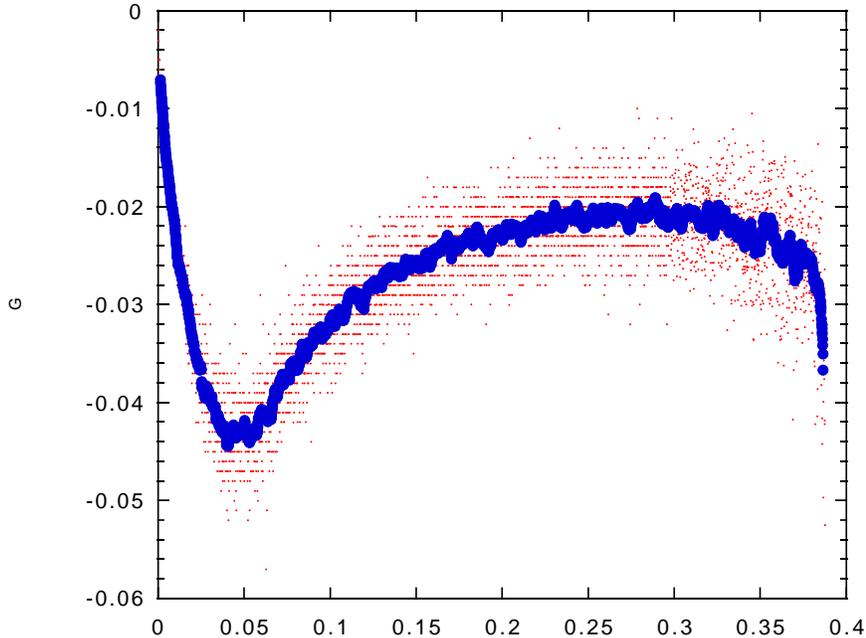}
      \end{center} \caption{ \label{CERT}
The quantities $1-s_{F}(i)$ (continuos curve) and the quantity $\Delta(i)$, averaged on a window of ten decimations 
(scattered points) of as function of the 
fraction of decimated nodes for one problem with $N=3\ 10^{5}$ and $\alpha=4.2$. The variable $i$ is the decimated 
node.}\end{figure}

In fig \ref{FINAL} we plot the complexity density $\Sigma_{M}/M$ ($M$ is the number of undecimated nodes) at the 
starting point and at the final point of the decimation procedure.  We see that the initial complexity density 
extrapolates to zero at $ \alpha\approx 4.267$ (in perfect agreement with the analytic estimates \cite{MPZ,MZ}.)
 while the final complexity 
becomes negative at $ \alpha_{A}\approx 4.252$. Similar results are obtained for  smaller values of $N$.

The conclusion is that in the present form the survey decimation algorithm may work in the infinite $N$ limit for $ 
\alpha <\alpha_{A}\approx 4.253<\alpha_{c}$.  The numerical experiments seem also to indicate that near $\alpha_{A}$ the 
complexity becomes negative near $f=1$.  The reasons for this remarkable phenomenon will not be discussed here.

Very similar results are obtained  if we use the certitude $s_{C}(i)$ to select the spins: there are very minor differences 
which need a very careful analysis to be evidenziated at least if we are not to near to $\alpha_{A}$, that may slightly 
depends on the method used. En passant we have also verified that the quantity $s_{C}(i)$ is strongly 
correlated with $\Delta(i)$ and the high order terms in $\epsilon$ are nor very important. In fig. (\ref{CERT}) we see  
for $N=3\ 10^{5}$ and $\alpha=4.2$ these two quantities  as function of $f$. It is remarkable that in the average these two 
quantities coincide, i.e. if we smooth $\Delta(i)$ on a sufficiently large window it becomes very near to $1-s_{F}(i)$ 

\begin{figure} \begin{center}    
      \includegraphics[width=0.70\textwidth]{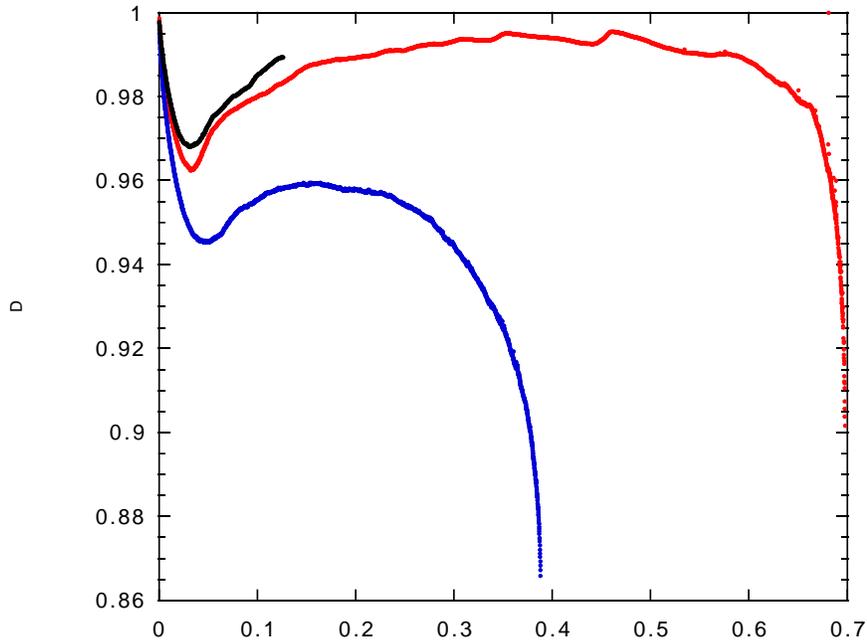}
      \end{center} \caption{ \label{POL}
 The polarization $|s_{T}-s_{F}|$ of as function of the 
fraction of decimated nodes for one problem with $N=3\ 10^{5}$ and $\alpha=4.2,\  4.25, \ 4.26$ from below to above.
}\end{figure}

The behaviour of the polarization (i.e. $|s_{T}-s_{F}|$) of the chosen variable as function of the fraction $f$ of 
removed variables is very similar to that of the certitude (the two quantities are strongly correlated) and it is shown 
in fig. (\ref {POL}).

The behaviour of both quantities is remarkable.  The behavior for small $f$ (e.g. $f<.02$) can be easily understood and 
it can be obtained from the distribution of the surveys of the undecimated problem.  The increase of the polarization of 
the chosen spin after the minimum around $f<\approx 05$ is an effect of computing the solution of the surveys equation 
in the decimated problem.  It is a very interesting phenomenon  and it is at the root of the good performances of the survey 
decimation algorithm.  

\begin{figure} \begin{center}    
      \includegraphics[width=0.70\textwidth]{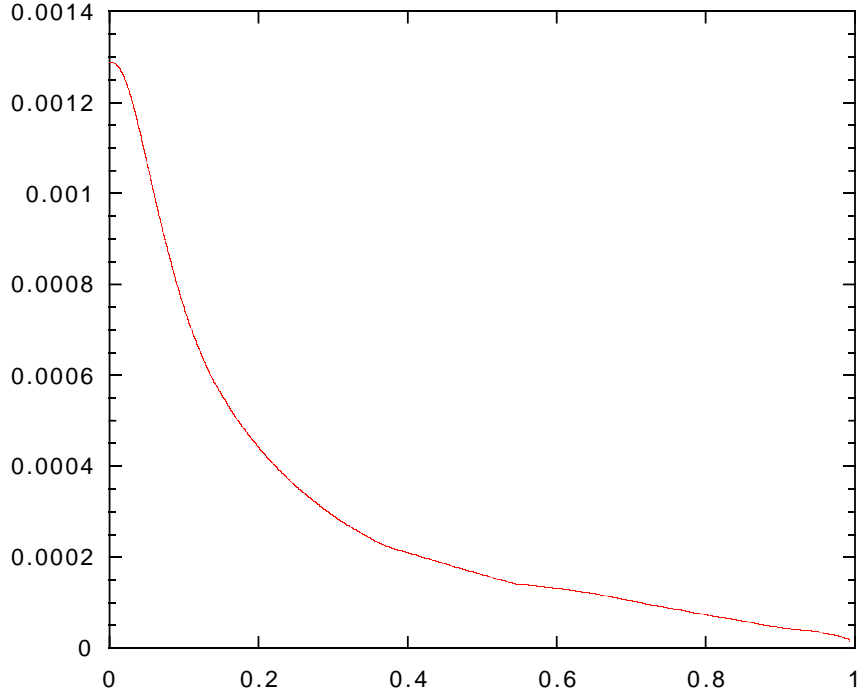}
      \end{center} \caption{ \label{CRIT}
 The complexity as function of the fraction $f$ 
fraction of decimated nodes for one problem with $N=3\ 10^{5}$ and $\alpha=4.2525$ The complexity jumps to zero at 
$f=.993$ where the number of undecimated clauses is about 2000.
}\end{figure}

The numerical experiments seem also to indicate that near $\alpha_{A}$ the value of $f$ where the complexity becomes 
negative goes to one, al least with the algorithm where the decimated clause has the maximal certitude.  In fig.  
(\ref{CRIT}) we see the complexity as function of $f$ for a sample with $N=3\ 10^{5}$ and $\alpha=4.2525$.  Here the 
complexity jumps to zero at $f=.993$.  However one should do a more careful and accurate  finite size analysis data to
see how this effect depends on the algorithm and on the sample.

Let us just sketch a simple intuitive argument for explaining this behaviour of the system.  Let us assume that:
\begin{itemize}
    \item The complexity can jump to zero when the non-trivial solution of the survey disappear only if the value of the 
    complexity is near to zero or negative.

    \item The probability for the decimation process to be stopped by the presence of a zero in the denominator of eq.  
    \ref{EQ} is  is small for small $\Sigma$ and it has a natural 
    prefactor that diverges when $f$ goes to one.
 
    \item At fixed $f$ the complexity is a decreasing function of $\alpha$:
$\partial \Sigma(f,\alpha)/ \partial \alpha< 0 $.

\end{itemize}
If the maximum value of $f$ would be less than one at $\alpha_{A}$, we would find a contradiction in the behaviour at 
$\alpha$ slightly greater that $\alpha_{A}$: the survey decimation process would end with a jump from a negative 
complexity and this is prohibited.  The contradiction would not be present if the maximum value of $f$ is 1, because the 
stopping probability diverges here.

In order to explain the performances of the algorithm it would important to find a more direct argument
that the maximum value of $f$ is 1 at $\alpha_{c}$.

\section{Conclusion}

The main result of this paper is the identification of the quantity (i.e. the certitude $1 - \min(s_{T},s_{F})$) that 
controls the complexity reduction during the decimation and the identification of the threshold value of $\alpha_{A}$ 
where the decimation algorithms must stop to work. Numerical simulations indicate that interesting phenomena happens 
near $\alpha_{A}$, however a more careful investigation is needed in order to properly quantify them. An analytic 
understanding of these phenomena is lacking at the present moment: it would be very important to obtain it because it 
would a key step in understanding the reasons for the good performances of the survey decimation algorithm.

\section*{Acknowledgements}
I thank Marc M\'ezard and Riccardo Zecchina for useful discussions and exchange of information.

\begin{small}

\end{small}

\end{document}